\documentstyle[aps,epsf,preprint]{revtex}
\newcommand{\pspicture}[1]{
\centerline{\setlength\epsfxsize{9.2cm}\epsfbox{#1}}}
\ifx\DeclareFontShape\undefined
\newcommand{\mathbf}[1]{{\bf #1}}
\newcommand{\mathrm}[1]{{\rm #1}}
\newcommand{\mathcal}[1]{{\cal #1}}
\else
	\DeclareSymbolFont{lasy}{U}{lasy}{m}{n}
	\SetSymbolFont{lasy}{bold}{U}{lasy}{b}{n}
	\let\Box\undefined
	\DeclareMathSymbol\Box{0}{lasy}{"32}
\fi

\ifx\epsffile\undefined
\else
\fi

\newcommand{\GeV}{\mbox{GeV}}

\newcommand{\plus}{\makebox[15pt][c]{$+$}}
\newcommand{\minus}{\makebox[15pt][c]{$-$}}

\newcommand{\er}[2]
{\hskip-0.5em\raisebox{0.08em}{\scriptsize{$\;\begin{array}{@{}l@{}}
\plus\makebox[0.25em][r]{#1\hfill} \\[-0.12em]
\minus\makebox[0.25em][r]{#2\hfill}
\end{array}$}}}

\newcommand{\err}[2]
{\hskip-0.5em\raisebox{0.08em}{\scriptsize{$\;\begin{array}{@{}l@{}}
\plus\makebox[0.50em][r]{#1\hfill} \\[-0.24em]
\minus\makebox[0.50em][r]{#2\hfill}
\end{array}$}}}

\newcommand{\AmS}{{\protect\the\textfont2
  A\kern-.1667em\lower.5ex\hbox{M}\kern-.125emS}}

\begin{document}

\begin{titlepage}

\begin{flushright}
Glasgow Preprint GUTPA 95--9--2 \\ Liverpool Preprint LTH 359 \\
hep-lat/9509083
\end{flushright}

%\vspace*{5mm}

\begin{center}
{\Huge Masses and decay constants of the light mesons in the quenched
approximation using the tadpole--improved SW-clover action.
\footnote{Presented at the {\rm XIII}th International Symposium
on Lattice Field Theory, Melbourne, Austrailia, July 11--15, 1995.}
}\\[10mm] {\it \large UKQCD Collaboration}\\[3mm]

{\bf C.~Michael}\\ {DAMTP, University of Liverpool, Liverpool, L69
3BX, U.K.}

{\bf H.~Shanahan}\\ Department of Physics and Astronomy, University of
Glasgow, Glasgow, G12 8QQ, Scotland, U.K.

\end{center}

\begin{abstract} We present results for the masses and decay constants
of the light mesons in quenched QCD using the standard gluon action
and a tadpole--improved SW--clover fermionic action to reduce
discretisation errors. The calculation has been carried out at fixed
volume and three lattice spacings corresponding to $\beta=5.7$, $6.0$
and $6.2$. We make comparisons with the conventional SW--clover
scheme. We use our results to extract continuum limits and to quantify
the size of discretisation errors at smaller $\beta$-values.
\end{abstract}

\vspace{5mm}

\vfill

\end{titlepage}

\maketitle

\section{INTRODUCTION}

To extract reliable continuum results from lattice studies, it is
essential to explore a range of $\beta$-values.  It is also valuable
to employ a fermionic discretisation which has minimal discretisation
errors.  Here we investigate the tadpole--improved clover--SW action
and we present preliminary results for the quenched spectrum at three
lattice spacings. We compare our light hadron results at $\beta=6.2$
with the conventional SW--clover approach and find no statistically
significant evidence of further improvement at that $\beta$-value.

We use non-local sources to allow a more accurate extraction of ground
state masses and couplings.  Combining our results at all three
$\beta$-values, we can extract the continuum limit and assess the size
of discretisation effects.  We find that dimensionless ratios of
observables have small discretisation corrections. We highlight the
uncertainties remaining in extracting continuum decay constants
because of the reliance on perturbative matching coefficients.

\section{THE SW CLOVER ACTION}

The SW fermion action is written in the form

 \begin{equation} S^C_F = S^W_F(\kappa) - \kappa c S_{\rm clover}
\end{equation} with hopping parameter $\kappa$. The quark fields $q$
and $\overline{q}$ which appear in any observable must also be
transformed (rotated). For an on-shell observable bilinear in the
quark fields, the equation of motion can be used to express this as an
improved observable given by~\cite{heatlie,zeds} \begin{equation}
O^{\rm imp}_{\Gamma} = (1+am_q(1-z))\bar{q} (\Gamma +z
\Gamma^{\otimes}) q
\label{eq:rotation}
 \end{equation} where $0 \le z \le 1$. The size of the clover term in
the action is determined by the coefficient $c$. At tree-level, $c$ is
equal to one.  As demonstrated by Heatlie {\it et al.} \cite{heatlie},
the terms of order $a$ are removed for this choice of $c$.

A mean field estimate of this coefficient was suggested by Lepage and
Mackenzie \cite{kermit}. In this approach, the poor agreement of
lattice perturbation theory with Monte Carlo simulations of
short--distance observables is explained by the presence of
non--physical ``tadpole'' diagrams.  One expects the power expansion
for the gauge link \begin{equation} U_\mu(x) = 1 + iagA_\mu(x) +
{{\cal{O}}(a^2 g^2)}
\label{eq:gauge-expand}\;\; ,
 \end{equation} to be valid for sufficiently small $a$. However,
tadpole diagrams introduce ultraviolet divergences which can partly
cancel the factors of $a^2$.  A solution to this is to rescale the
gauge link by a factor $u_0$, such that $U_\mu(x) \rightarrow
\tilde{U}_\mu(x) = {U_\mu(x)}/{u_0}$ .  The factor $u_0$ is chosen so
that a Monte Carlo calculation of a short distance observable using
the rescaled gauge field agrees with the perturbative calculation.  An
appropriate choice is
\begin{equation}
u_0 = \langle \frac{1}{3} {\rm Tr} U_{\rm plaq} \rangle^{1/4} \;\; .
\end{equation}
 Inserting this factor into the SW action changes the hopping
parameter to $\tilde{\kappa}=\kappa u_0$ and the clover coefficient
$c$ from 1 to $u_0^{-3}$.  % \begin{equation}
%c = \langle \frac{1}{3} Tr U_{plaq} \rangle^{-3/4} \;\; .
% \end{equation}
 A perturbative evaluation of $c$ by Naik \cite{naik} gives a
numerically similar value which is some support for the assumption of
dominance of the tadpole diagrams.

\subsection{Computational Details}

Gauge configurations were calculated at $\beta=5.7$, $6.0$ and $6.2$,
using spatial volumes that are approximately the same. At each lattice
spacing, propagators at 2 quark masses, straddling the strange quark
mass, were calculated for each configuration --- see table 1.
Hadronic correlations were evaluated for each combination of such
quarks, yielding three mesonic masses. Hence the extrapolation to the
chiral limit could be explored using three hadronic masses.  The
configuration and propagators at $\beta=6.0$ and $6.2$ were calculated
on a 320 node Cray T3D.  The gauge configurations and propagators at
$\beta=5.7$ were calculated respectively on a 16 and 64 node i860
Meiko Computing Surface. For the transformation in
eq.~\ref{eq:rotation}, we take $z=0$ which corresponds to no rotation
in the propagators.

\begin{figure}
\pspicture{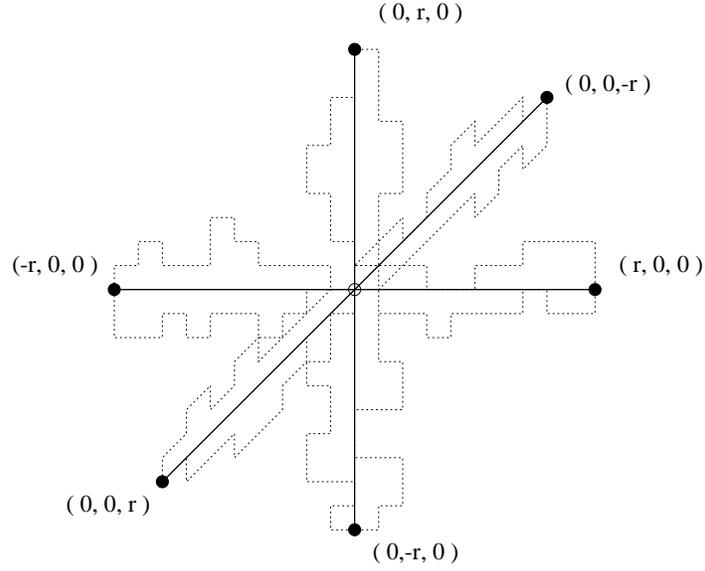}
\vspace{10pt}
\caption{Schematic guide to fuzzing. The source is composed of six points
around the origin. The dotted lines indicate typical paths that will
be used in the average.}
\label{fig:fuzzed}
\end{figure}

\begin{figure}
\pspicture{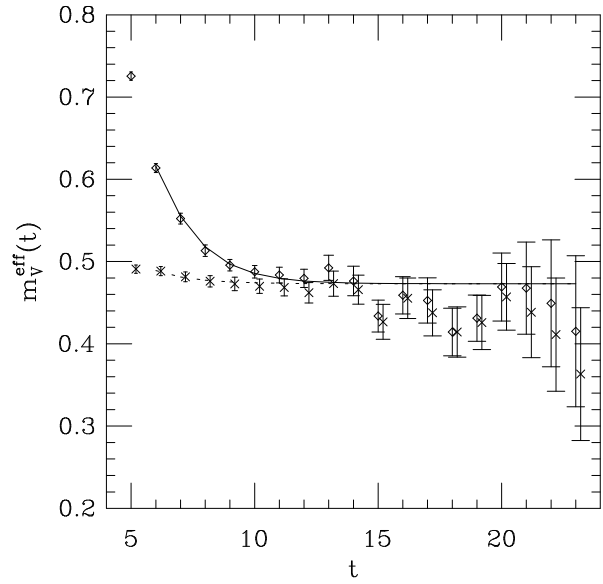}
\caption{The effective mass of the vector meson versus time
separation $t$. The data are at $\beta=6.0$ and $\kappa=0.1381$, using
local (diamonds) and fuzzed (crosses) sources. The fits shown are two
state expressions.}
\label{fig:eff-mass-fuzzed}
\end{figure}

The propagators were calculated using both non-local and local sources
so that ground state and excited state contributions to correlators
could be separated effectively. Local sinks were employed. At
$\beta=5.7$, Jacobi smeared sources \cite{smearing} were used, and
correlations were measured with three combinations (neither, one,
both) of nonlocal and local quark propagators in a meson.  At
$\beta=6.0$, local and ``fuzzed'' sources were used
\cite{liverpool:fuzzing}. A fuzzed source can be written in the form
\begin{equation} S_{\rm fuzzed}(x) =
\sum_{i=1}^3 \sum_{y_i=\pm r} F(y) \delta(x-y_i) \;\; ,
 \end{equation} where $F(y)$ is an SU(3) matrix representing the
average of the fuzzed paths of gauge links from the origin to $(y)$
(see fig.~\ref{fig:fuzzed} for a schematic description). Fuzzing is
less computer expensive than Jacobi smearing, as the only numerical
work involves the averaging of the gauge links. It is gauge invariant
and avoids time spent in gauge fixing. We use the fuzzing prescription
with 5 iterations as in ref.~\cite{liverpool:fuzzing} and we choose
$r=6a$ at $\beta=6.0$. An example of the advantage of fuzzing at the
source is shown in fig.~\ref{fig:eff-mass-fuzzed} where an efficient
extraction of the ground state vector meson is seen.

\section{FITTING}

 We focus on the pseudoscalar and vector mesons. In order to evaluate
accurately the central value and statistical errors, we fit to as many
channels of data as possible. Furthermore, to isolate the ground state
for a wide range of time slices, we use a two-exponential fit.

The two--point correlations can be expressed as a sum over transfer
matrix eigenstates \begin{equation}
\sum_{\vec{x}} \langle 0 | O_2( \vec{x}, t ) O^\dagger_1(0) | 0 \rangle
 = \sum_i {c_2^i c_1^{i*} \over m_i} \cosh{m_i(t-N_t/2)} \; .
\end{equation}
Here $c^i = {\langle 0|O|H^i\rangle }$ where $O_{1,2}$ are operators
that have overlap with the ground and excited mesonic states
$|H^0\rangle$ and $|H^1\rangle$. These operators can be local or
nonlocal, or have different gamma matrix structures. For both mesons,
we fit simultaneously to all the available local and non-local source
correlations of two--point functions.

In the case of the pseudoscalar, we note that the fields $P(x) =
\overline{q}(x) \gamma^5 q(x)$ and  $A(x) = \overline{q}(x) \gamma^0
\gamma^5 q(x)$ will create/annihilate the pseudoscalar, hence a
simultaneous fit is performed to two--point functions with the four
combinations PP, AP, PA, AA. These contributions can be described in a
factorising fit by the amplitudes $\langle0|A|PS\rangle$ and
$\langle0|P|PS\rangle$, where $|PS\rangle$ is the pseudoscalar state.
This provides tight constraints on the fit.

\begin{figure}
\pspicture{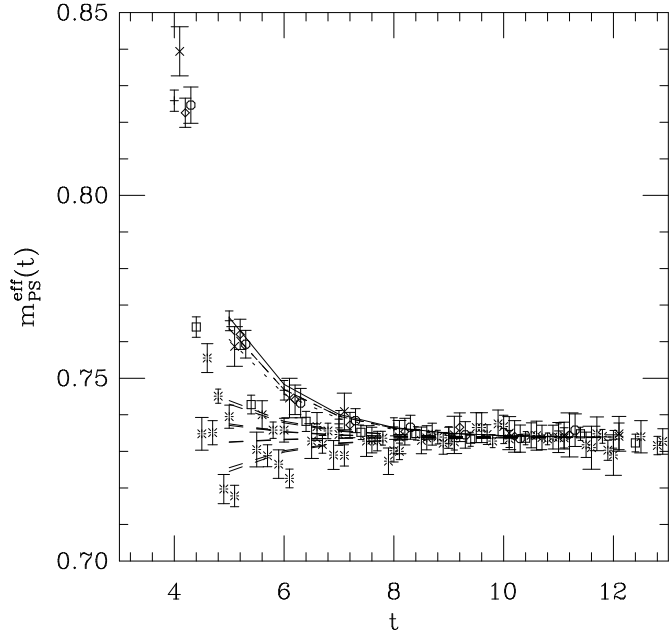}
\caption{Effective mass plot from the local and smeared combinations of
the operators $A$ and $P$ at $\beta=5.7$ and $\kappa=0.13843$.}
\label{fig:effective_mass_ps}
\end{figure}

We define the lattice decay constant for a pseudoscalar meson,
$\hat{f}_{PS}$ as : \begin{equation}
\langle 0 | A_{L} | PS \rangle =  m_{PS} \hat{f}_{PS} \;\; ,
\label{eqn:f_ps-def}
 \end{equation} where $A_L$ is the {\it local} axial current operator.
The lattice $\hat{f}$ will be related to the continuum value $f$ by
including quark field rotations, setting the scale $a$ and including
the appropriate matching coefficient $Z_A$.  One can determine
$\hat{f}_{PS}$ directly from the factorising fit, without fitting to a
ratio of channels as was used previously \cite{ukqcd:light-hadrons}.
An example of the result of such fits is shown in
fig.~\ref{fig:effective_mass_ps}.

Likewise, simultaneous fits for the vector meson can be performed
using the operators $V_1^i = \overline{q} \gamma^i q$ and $V_2^i =
\overline{q} \gamma^0 \gamma^i q$, and different  smearing/fuzzing
combinations.  The lattice vector decay constant, $\hat{f}_V$ is
defined by \begin{equation}
\sum_{j=1}^3 \sum_{\vec{x}} \langle V_1^j(\vec{x},t)
{V_1^{j}}^\dagger(0) \rangle
\rightarrow
%\frac{3 m_V^2}{ 2 \hat{f}_V^2 } e^{-m_V t}
%& + & A^*_V e^{-m_V^* t}
\frac{3 m_V^2}{ \hat{f}_V^2 } \cosh{m_V(t-N_t/2)}
 + A^*_V \cosh{m_V^*(t-N_t/2)}
\; ,
\label{eqn:twopt-vector}
 \end{equation}
%(plus terms arising from  $(t \leftrightarrow N_t-t)$),
where $V_1^j$ is the local vector current.

Despite the quite large statistics used in this calculation, the large
number of different types of two-point functions requires that a full
correlated fit must be carried out carefully. In particular, we employ
the methods described by Michael and McKerrell\cite{liverpool:fitting}
for the inversion of the correlation matrix. At $\beta=6.0$ and $6.2$,
we model the correlation matrix with two exponents using the
5-diagonal approximation of ref.~\cite{liverpool:fitting}. At
$\beta=5.7$, we average all but 20 of the largest eigenvalues of the
correlation matrix.  We fit to the largest range of $t$ consistent
with an acceptable $\chi^2$. The errors quoted in this preliminary
report are statistical only.

\section{RESULTS: MASSES}

\subsection{Chiral Extrapolation}

The ground state pseudoscalar and vector masses are determined by the
fits to two-particle correlations as described above. The issue of the
rotations of the quark fields is irrelevant to mass determinations and
will be discussed later. We define $\kappa_{\rm crit}(\beta)$ by the
requirement $m_{PS}^2 = 0$ for each lattice. At each lattice spacing,
a linear extrapolation in the two squared pseudoscalar masses (where
the quark masses are degenerate) was carried out.  The results for
$\tilde{\kappa}$ are given in table 2. A full fit using also the meson
mass determined from the non--degenerate quark--mass case is possible,
however, the fits gave statistically significant evidence for some
curvature in $m_{PS}^2$ versus $1/\kappa$.

Without tadpole--improvement, there has been no agreement of
non-perturbative lattice determinations of $\kappa_{\rm crit}(\beta)$
with one loop perturbation theory. For Wilson-like fermion actions, it
is found that the tadpole--improvement factor of $u_0$ makes
$\tilde{\kappa}_{\rm crit}$ closer to the tree-level value of 1/8.
For example, a determination of $\kappa_{\rm crit}$ using the $c=1$
clover action~\cite{ukqcd:light-hadrons} at $\beta=6.2$ gave a result
of $\kappa_{\rm crit}=0.14315\pm 2$ which corresponds to
$\tilde{\kappa}_{\rm crit} = 0.12670\pm{2}$.

For the tadpole--improved action explored here, we present in table~2
the comparison of the non-perturbative results for
$\tilde{\kappa}_{\rm crit}$ with the one loop perturbative
calculation. Here we have chosen to use a coupling derived from
$\alpha=-{3 \over 4 \pi} \log S_{\rm plaq}$. Although the agreement is
excellent at $\beta=6.2$, the trend as $\beta$ is decreased is not.
This lack of agreement of $\kappa_{\rm crit}$ between non-perturbative
determinations and one--loop perturbation theory at smaller $\beta$
will cause some uncertainty in normalisation when quark mass factors
are included in the determination of the decay constants.

\subsection{$m_V^2 - m_{PS}^2$}
 The experimental value of the difference $m_V^2 - m_{PS}^2$ remains
remarkably constant, at approximately $0.55 \, \GeV^2$, over a wide
range of masses.  Quenched lattice simulations of relativistic quarks
have been in poor agreement with this result, and the discrepancy
increases as the quark mass is increased. In the heavy quark regime,
such splitting has a leading order dependence on the chromo--magnetic
coupling of the heavy quark with the gluon field. This is precisely
the term which is amplified in the tadpole--improved clover action.
While one cannot make a similar statement in the light quark regime,
it is plausible that this modification of the clover term will also
have a significant effect there.

\begin{figure}
\pspicture{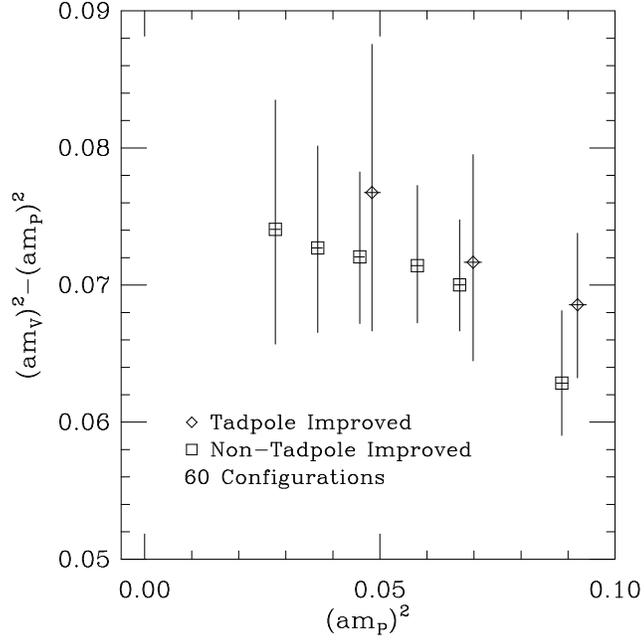}
\caption{The comparison of the mass combination $(am_V)^2-(am_P)^2$
at $\beta=6.2$ from SW-clover and tadpole-improved SW-clover fermionic
actions.}
\label{fig:tadpole-comparison}
\end{figure}

A comparison at $\beta=6.2$ can be made with non tadpole-improved
UKQCD data~\cite{ukqcd:light-hadrons,UKQCD:J}, based on 60
configurations.  These configurations exist as a subset of the $184$
configurations available now. A comparison of the results is shown in
fig.
\ref{fig:tadpole-comparison}. Unfortunately, there is no significant
statistical difference between the two actions. We will soon have
results for fuzzed sources in the tadpole-improved mass extraction
which should lead to a reduction of errors.

\begin{figure}
\pspicture{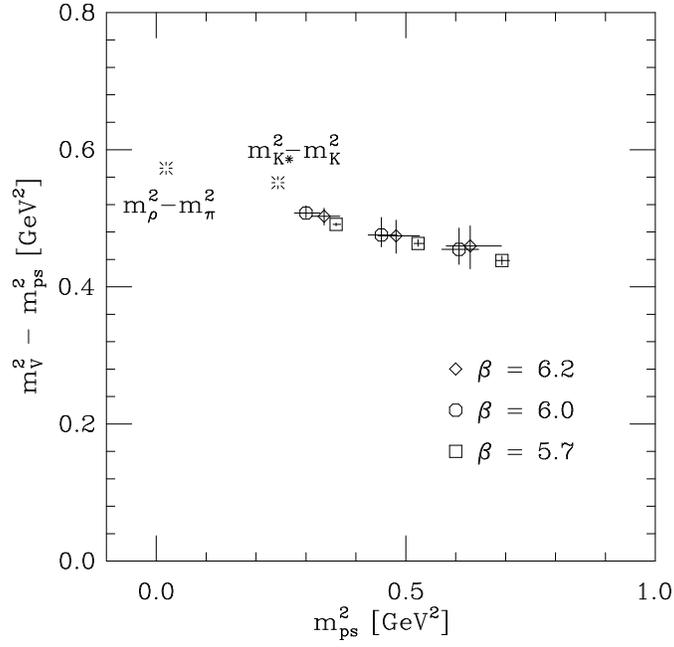}
\caption{The comparison of the mass combination $(m_V)^2-(m_P)^2$
between lattice results and experiment.}
\label{fig:mvsq-minus-mpssq}
\end{figure}

The behaviour at all three lattice spacings is shown in
fig.~\ref{fig:mvsq-minus-mpssq}. The scale ($a^{-1}_\rho$) is
determined from the chirally extrapolated $m_\rho$. The fall-off of
the splitting with respect to $m_{PS}^2$ is slightly faster than one
would expect by comparison with the physical splittings
$m_\rho^2-m_\pi^2$ and $m_{K*}^2-m_K^2$. However, until this
difference is examined for quark masses around that of the charm
quark, the advantage of tadpole improvement in this case remains
unproven.

\subsection{Discretisation effects}

The dimensionless ratio of masses is known to have discretisation
errors of order $\alpha a$ for the SW-clover fermionic action.  The
tadpole-improved variant should preserve this result and is expected
to improve on it by reducing the coefficient. We explore this by
studying such ratios of observables against an observable proportional
(to leading order) to $a$.

\begin{figure}
\pspicture{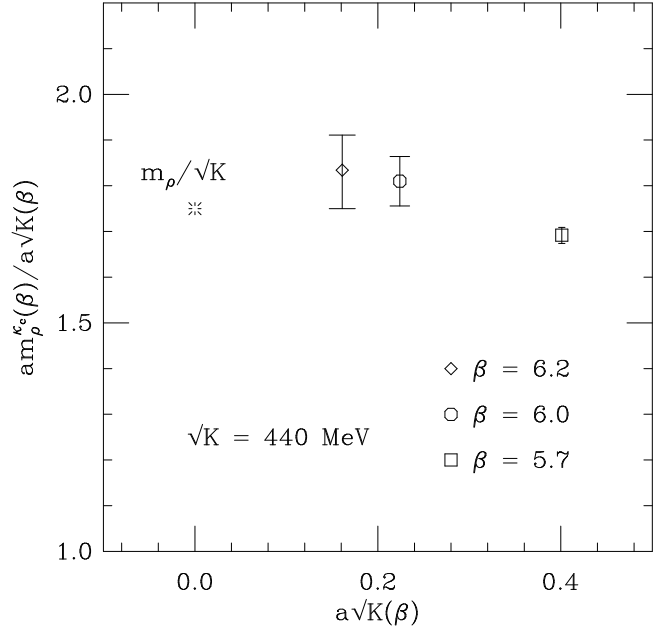}
 \caption{ The dimensionless ratio of the $\rho$-mass to the square
root of the string tension. The ``experimental'' point at $a=0$
corresponds to (770MeV/440MeV).}
\label{fig:mrho-on-tension}
\end{figure}

Since the Wilson gauge action has inherent discretisation errors of
order $a^2$, it is advantageous to use this as a reference. We present
ratios of $m_{\rho}$ to $\sqrt{K}$ where $K$ is the string tension
determined by interpolating all existing data. In
fig.~\ref{fig:mrho-on-tension}, we find that the ratio is constant
(within statistical errors) at $\beta=6.0$ and $\beta=6.2$.  There is
however, a significant deviation from this constant for the ratio at
$\beta=5.7$.  Since this behaviour is consistent with $a^2$, it could
be ascribed to the discretisation errors in either (or both) of the
pure-gauge or fermionic quantities.

 It is important to compare lattice results in the quenched
approximation without chiral extrapolation if possible, since there
are known to be anomalies in the chiral limit of a quenched theory. As
suggested by Lacock and Michael \cite{UKQCD:J}, a dimensionless
quantity can be constructed using this data without having to perform
a chiral extrapolation: \begin{equation} J = m_V {d m_V \over d
m_{PS}^2} \end{equation} where the slope and $m_V$ are to be evaluated
at the vector mass satisfying $m_V/m_{PS}=1.8$ which corresponds to
the $K^*$ meson.

\begin{figure}
\pspicture{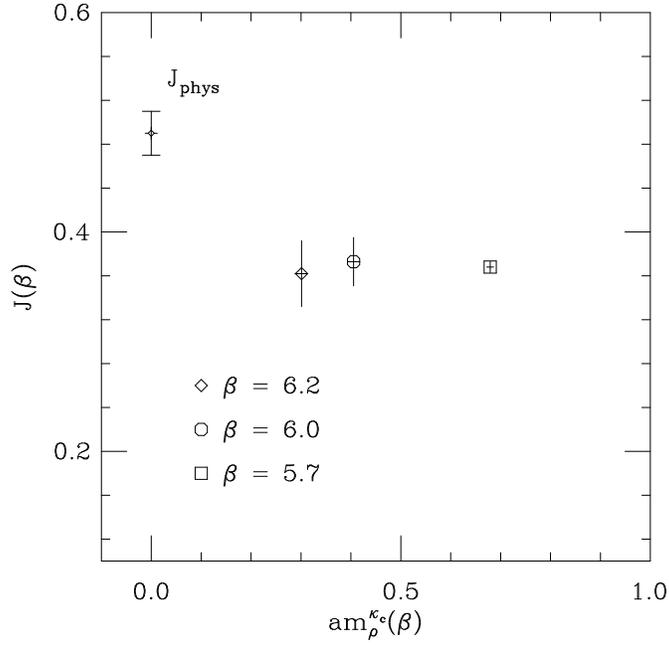}
\caption{ The  dimensionless quantity $J = m_v \, d m_V / d m_{PS}^2$
from the different lattice $\beta$-values compared to the experimental
value. }
\label{fig:J}
\end{figure}

Our results are shown in fig.~\ref{fig:J} and there is no evidence of
any discretisation error since the values of $J$ from each lattice
spacing are equal within errors. Our result agrees with that from most
existing quenched lattice data which are consistent~\cite{UKQCD:J}
with a value of $J$ of 0.37. This value is in disagreement with
experimental data and, to substantiate this discrepancy, we will need
a careful study of systematic errors, particularly at our largest
$\beta$-value.

\section{RESULTS: DECAY CONSTANTS}

As can be seen from eqs.~\ref{eqn:f_ps-def} and
\ref{eqn:twopt-vector}, the parameters $\hat{f}_{PS}$ and
$1/{\hat{f}_V }$ can be easily obtained from the lattice data.
However, to evaluate the physical quantities $f_{PS}$ and $1/f_V$, one
has to specify a suitable normalisation for the quark propagators and
a renormalisation factor for the lattice operators. For fermion
actions of the kind of eq.~1, several equivalent solutions have been
proposed.  To estimate the size of the systematic error induced by
relying on perturbative estimates of the matching factors, we present
here two interpretations of the action (eq.~1) with $c=u_0^{-3}$ which
result in relatively large differences.

This systematic uncertainty is due to both our ignorance on the
contribution of $O(\alpha^2_s)$ terms in the perturbative expansion,
and to the possible different interpretations of the tadpole
improvement procedure.

\noindent
{\bf Normalisation I}

In this case we interpret the clover coefficient $c=u_0^{-3}$ as a
consequence of the tadpole improvement, which implies a redefined
hopping parameter ($\tilde{\kappa}$) and a gauge field rescaling.
Because $U_{\mu}\to U_{\mu}/u_0$, the covariant derivative (or the
lattice bare mass for the unrotated case $z=0$ which we are using)
which appears in eq.~2 has to be divided by $u_0$.  We choose,
however, to adopt a slightly different normalisation, which amounts to
using the renormalized quark mass, \begin{equation} am_q \;\; = \;\;
\frac{1}{2}\left( \frac{1}{\tilde{\kappa}} -
\frac{1}{\tilde{\kappa}_c} \right) \; ,
\label{eqn:pole-mass}
 \end{equation} instead of the bare mass since the ``rotation'' factor
$1+am_q$ is then relatively closer to unity and possibly better
reproduced by perturbation theory. This is possible because formally
the difference between the bare mass and the renormalised mass only
contributes at order $\alpha a$ which does not effect the improvement.

The renormalisation constants to be used in this case differ from
those corresponding to the SW action, and have been computed in
ref.~\cite{ns}, and are \begin{eqnarray} Z_A \;\; & = \;\; 1 - 0.41
\alpha_s + {\cal O}(\alpha_s^2) \; , \nonumber \\ Z_V \;\; & = \;\; 1
- 0.58 \alpha_s + {\cal O}(\alpha_s^2) \; , \label{eqn:per-Z}
\end{eqnarray} where we choose to use $\alpha_s = -{3 \over 4 \pi}
\log S_{\rm plaq}$ as an appropriate improved estimate of the
coupling. There is an inherent systematic error from the choice of
scale for this coupling - this can only be resolved by a two-loop
calculation.

\noindent
{\bf Normalisation II}

Let us consider the results one obtains using the action of eq.~1 with
$c=u_0^{-3}$ but without attributing to such a coefficient the meaning
of tadpole subtraction. Because of that both the hopping parameter and
the normalisation of the propagator itself keep their original values.
Moreover, since the perturbative expansion of the clover coefficient
is 1, up to terms of $O(\alpha)$, the renormalisation constants are
the same as those calculated for $c=1$, given in ref.~\cite{zeds}.  In
this case we used the ``unrotated'' operators corresponding to taking
$z=0$ in eq.~2, where $m_q$ is now interpreted as the ``bare'' lattice
quark mass.

\begin{figure}
\pspicture{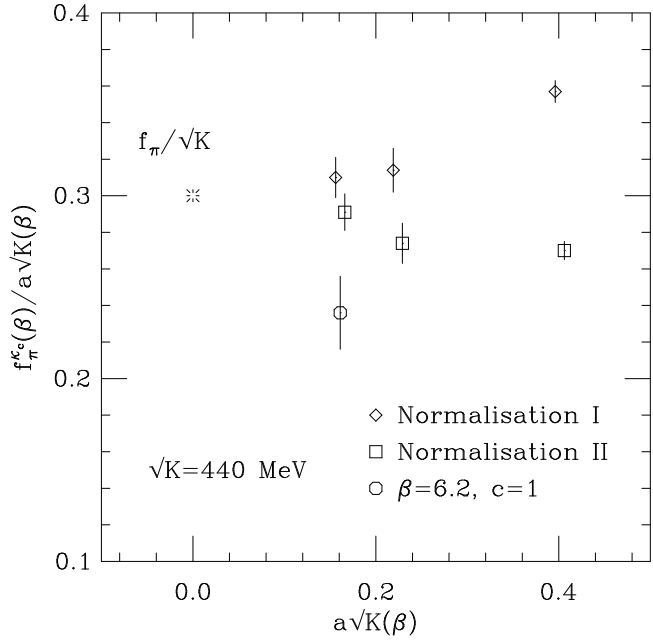}
 \caption{ The dimensionless ratio of $f_{\pi}$ to the square root of
the string tension. The ``experimental'' point at $a=0$ corresponds to
(132MeV/440MeV).
%The different schemes are as discussed in the text
%with I being a tadpole-improved approach and II a conventional clover
%interpretation.
The points are displaced slightly for clarity.}
\label{fig:f-pi-on-string}
\end{figure}

\begin{figure}
\pspicture{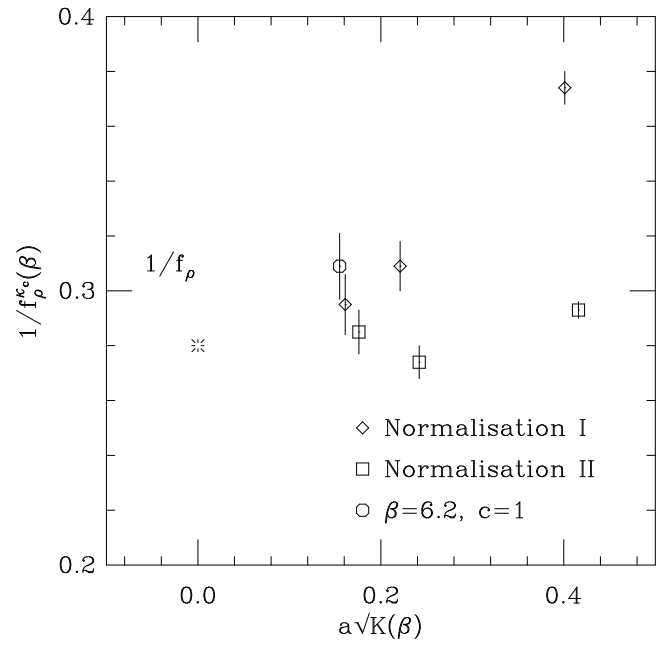}
\caption{The $\rho$ meson decay constant $1/f_{\rho}$ versus the
lattice spacing $a$. The points are displaced slightly for clarity. }
\label{fig:one-on-f-rho}
\end{figure}

In order to compare with other work, at each $\beta$, we extrapolate
the lattice decay constants to the chiral limit assuming a linear
behaviour with $m_{PS}^2$.  The dimensionless quantities
$f_{\pi}/\sqrt{K}$ and $1/f_{\rho}$ are shown for each normalisation
in figs.~\ref{fig:f-pi-on-string}, \ref{fig:one-on-f-rho}.

The results for $f_\pi/\sqrt{K}$ for both normalisation schemes are
consistent with having a common continuum limit which agrees with the
experimental value.  As $\beta$ is decreased, the difference between
normalisations I and II increases. This largely arises from the fact
that the perturbative estimate of $m_q$ is not in agreement with the
numerical value (it is therefore due to large unknown contributions at
$O(\alpha^2)$), and this causes a poor renormalisation in the case II.
An insight into this problem could be achieved by calculating the
``rotated'' operators of eq.~2 with $z=1$, which would permit us to
eliminate the uncertainty in the renormalisation of the quark mass.
Such a study is currently in progress. Ideally, however, one should
measure the matching constants non-perturbatively, as outlined in
refs.~\cite{ward} and \cite{APE:renorm}.

The $c=1$ results~\cite{UKQCD:J} for $f_\pi/\sqrt{K}$ at $\beta=6.2$
are seen to lie significantly lower in fig.~\ref{fig:f-pi-on-string}.
A careful study of systematic errors is needed to explore this
discrepancy further. Since we have used different rotation factors
($z=0$ for $c=u_0^{-3}$ and $z=1$ for $c=1$), the effect of this
difference on the discrepancy can also be determined by a study of
$z=1$ for the tadpole--improved case.

Fig.~\ref{fig:one-on-f-rho} shows for $1/f_\rho$ a rather similar
picture of agreement between the continuum limit of the
tadpole--improved schemes and experiment. In this case the $c=1$
result is also in agreement.

\section{CONCLUSIONS}

The goal of lattice investigations of QCD is to find a formalism which
has small discretisation errors. This has to be studied by obtaining
results for a wide range of observables for several lattice spacings.
In this paper we have outlined preliminary results on the spectrum and
couplings from a tadpole--improved SW clover action.  We find that,
within our present statistical errors, the results at $\beta=6.0$ and
6.2 are consistent while those at $\beta=5.7$ show signs of
discretisation effects. This is encouraging since it suggests that
meaningful results can be obtained at 6.0 which corresponds to $a^{-1}
> 1.8 \, \GeV$. To push the frontier down to 5.7, it seems plausible
that an improved gauge action will also be needed since the ratio of
$m(0^{++})/\sqrt{K}$ evaluated using the Wilson gauge action is known
to deviate significantly at $\beta=5.7$ so that there are significant
order $a^2$ discretisation errors here. % \cite{edwards}.

We also emphasize that continuum extraction of decay constants and
matrix elements is beset with ambiguity if a perturbative matching is
used.  This arises in part because the critical hopping parameter is
not accurately given by 1 loop perturbation theory. A non-perturbative
determination of these matching coefficients is essential for fully
reliable continuum predictions.

\begin{center}
\begin{table}
\begin{tabular}{|c|c|c|c|c|c|}
\hline
$\beta$ & $N_s^3 \times N_t$ & Number of & $\kappa$ & Smearing &
Clover \\ 	&	& 	configurations & & & coefficient \\
\hline 5.7 & $ 12^3 \times 24$ & 480 & 0.13843, 0.14077 & Jacobi \&
Local & 1.5678 \\ \hline 6.0 & $ 16^3 \times 48$ & 130 & 0.1370,
0.1381 & Fuzzed \& Local & 1.4785 \\ \hline 6.2 & $ 24^3 \times 48$ &
184 & 0.1364, 0.1371 & Local & 1.4424 \\
\hline
\end{tabular}
\vspace{10pt}
\caption{Input parameters for simulation.}
\end{table}
\end{center}

\begin{center}
\begin{table}
%\label{tab:kappa-crit}
\begin{tabular}{| c | c | c | c | c |}
\hline
$\beta$ & $\tilde{\kappa}_{\rm crit}$ & $\tilde{\kappa}_{\rm
crit}^{\rm 1 Loop}$ & $a m_\rho(\kappa_{\rm crit})$ & $a^{-1}_{m
\rho}$ [GeV] \\ \hline 5.7 & 0.12344\er{2}{2} & 0.1214 &
0.678\er{7}{7} & 1.14\er{1}{1} \\ \hline 6.0 & 0.12218\er{3}{2} &
0.1219 & 0.405\err{13}{12} & 1.90\er{6}{6} \\ \hline 6.2 &
0.12206\er{1}{2} & 0.1221 & 0.302\err{12}{14} & 2.55\err{12}{10}\\
\hline \end{tabular}
\vspace{10pt}
\caption{Output results of simulation. Non-perturbatively derived values of
$\tilde{\kappa}_{\rm crit}$ compared with a 1--loop calculation using
tadpole--improved perturbation theory. The vector masses extrapolated
to the chiral limit ($m_\pi = 0$) and the resulting scale, taking
$m_\rho = 0.7699 \; \GeV$.}
\end{table}
\end{center}

\end{document}